\documentclass[aps,prl,twocolumn,unsortedaddress]{revtex4}

\usepackage{graphicx}
\begin{document}

\title{Origin of the structural phase transition in Li$_7$La$_3$Zr$_2$O$_{12}$}

\date{\today}

\author{N. Bernstein}
\author{M.~D. Johannes}
\affiliation{Center for Computational Materials Science, Naval Research Laboratory, Washington, DC 20375}
\author{Khang Hoang}
\altaffiliation{Resident at the Center for Computational Materials Science, Naval Research Laboratory, Washington, DC 20375.}
\affiliation{Computational Materials Science Center, George Mason University, Fairfax, VA 22030}

\begin{abstract} 

Garnet-type Li$_7$La$_3$Zr$_2$O$_{12}$ (LLZO) is a solid electrolyte material with a low-conductivity tetragonal and a high-conductivity cubic phase. Using density-functional theory and variable cell shape molecular dynamics simulations, we show that the tetragonal phase stability is dependent on a simultaneous ordering of the Li ions on the Li sublattice and a volume-preserving tetragonal distortion that relieves internal structural strain. Supervalent doping introduces vacancies into the Li sublattice, increasing the overall entropy and reducing the free energy gain from ordering, eventually stabilizing the cubic phase. We show that the critical temperature for cubic phase stability is lowered as Li vacancy concentration (dopant level) is raised and that an activated hop of Li ions from one crystallographic site to another always accompanies the transition. By identifying the relevant mechanism and critical concentrations for achieving the high conductivity phase, this work shows how targeted synthesis could be used to improve electrolytic performance.

\end{abstract}

\maketitle

The garnet structure material Li$_7$La$_3$Zr$_2$O$_{12}$ (LLZO) has an ionic conductivity that varies by two orders of magnitude depending on whether synthesis produces a cubic ($\sigma_{\rm cubic}$=1.9$\times$10$^{-4}$ S/cm) \cite{Murugan:2007eg} or tetragonal ($\sigma_{\rm tetra}$=1.63$\times$10$^{-6}$ S/cm) \cite{Awaka:2009jv} phase. With the higher conductivity, LLZO has excellent solid electrolyte material characteristics. Its stability against both Li metal and standard cathode LiCoO$_2$ \cite{Ohta:2011ie}, combined with a $\sim$5 eV band gap \cite{Murugan:2007eg} and high ionic conductivity, make it suitable for exploiting the full voltage difference between anode and cathode while circumventing the safety concerns inherent to all current liquid electrolytes. Initially, the process by which the cubic ground state could be stabilized against the tetragonal was unknown and uncontrolled. The relevant factor was eventually traced down to the addition of Al in the structure, whether via accidental uptake from Al-containing crucibles \cite{Geiger:2011cg} or via intentional doping \cite{Jin:2011wg,Allen:2012gk}. Inclusion of other supervalent ions, including Ta, Nb, and Ga \cite{Allen:2012gk,Adams:2011ho}, has also been successful in producing the high conductivity phase. However, the underlying mechanism that controls the transition has so far remained a mystery, hindering further progress toward improving the material for practical usage.

Here we use our recently developed variable cell shape density-functional theory (DFT) plus molecular dynamics (MD) method to investigate the driving force behind the tetragonal to cubic transition which subsequently raises the conductivity. We find that in the cubic phase, the Li sublattice is always disordered (all Li symmetry sites partially occupied), while in the tetragonal phase it is always ordered (all Li sites either full or empty). We find that the energy is lowered by a simultaneous ordering of the Li atoms that relieves Li$-$Li Coulomb repulsion but unfavorably distorts the ZrO$_6$ octahedra and a lattice distortion that restores the preferred Zr$-$O bonds. The two symmetry-breaking but volume-preserving phenomena always occur in conjunction and either one alone actually raises the total energy of the system. When Al$^{3+}$ is doped into the system, charge compensation takes place through creation of Li$^+$ vacancies that reduce the free energy advantage of complete ordering on the Li sublattice, eventually leading to disorder and a transition to cubic symmetry. The transition is always signaled by a sudden shift of Li occupation which can be used to pinpoint the critical temperature and vacancy concentration. Using these criteria, we estimate that the critical concentration of Al dopants necessary to achieve the high-conductivity cubic phase of Li$_{7-2x}$Al$_{x}$La$_3$Zr$_2$O$_{12}$ is $x=0.2$ (vacancy number $n_{\rm vac} = 2x = 0.4$ per formula unit) at zero temperature, in good agreement with experiment. We show that the cubic phase can be reached at {\it some} temperature, regardless of Li content, but that the critical temperature drops as a function of vacancy number. The understanding uncovered in this work will be useful for choosing more efficient dopants and further raising the ionic conductivity.

\begin{figure} 
\centerline{\includegraphics[width=0.5\columnwidth]{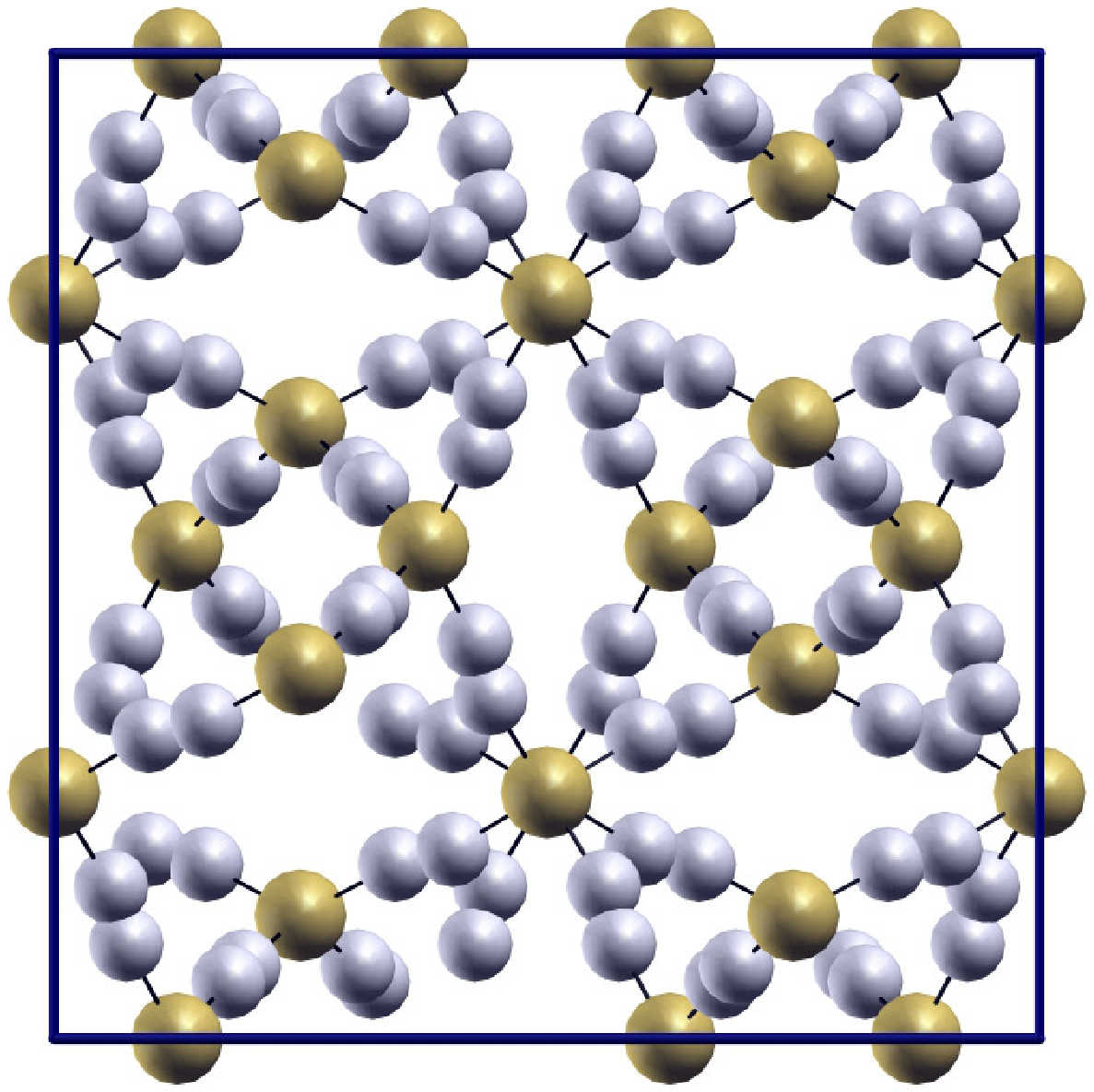} \includegraphics[width=0.5\columnwidth]{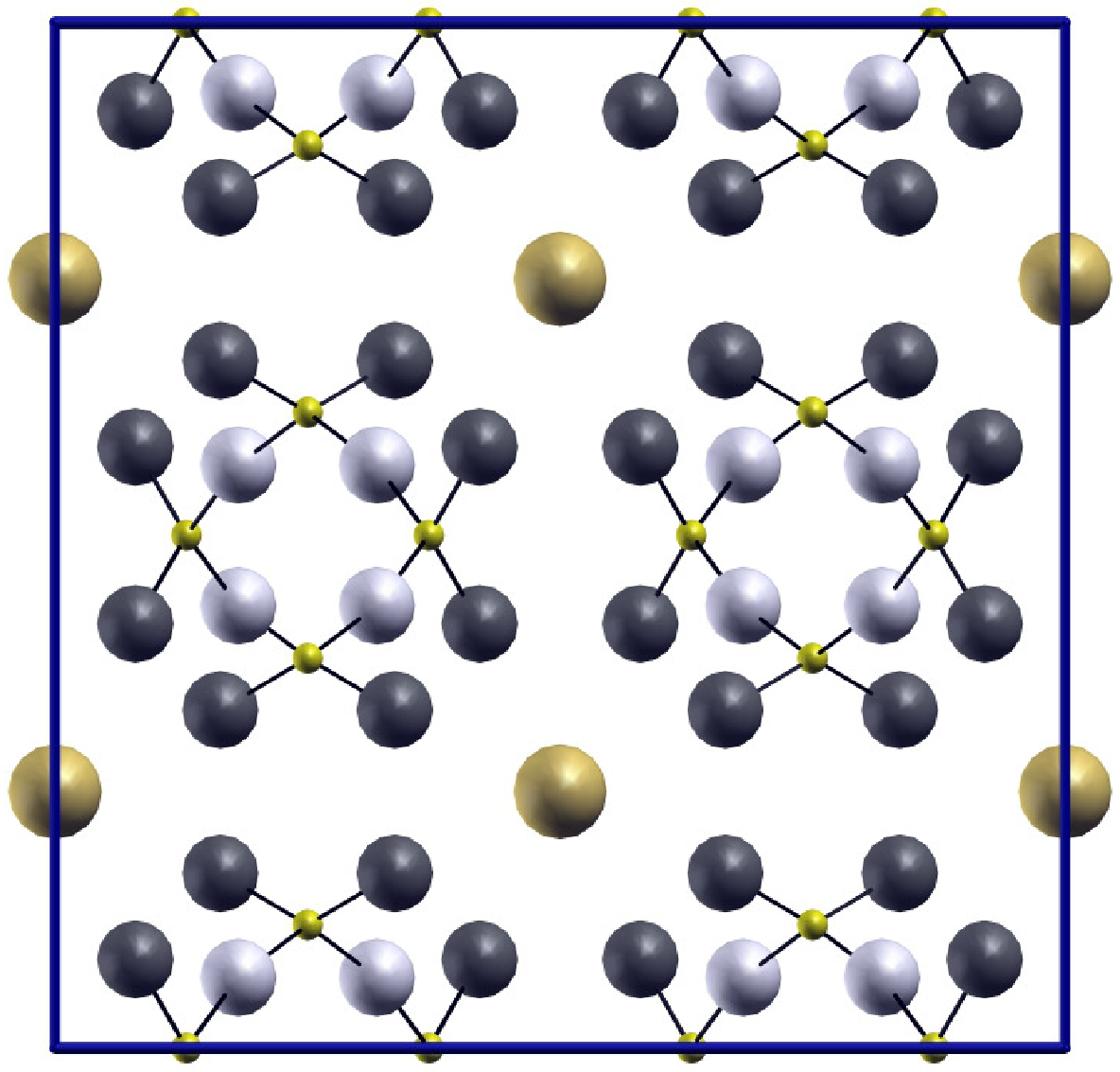} }
\caption{(Color online) Li sublattice in the cubic (left) and tetragonal (right) phases of LLZO. All Li positions are included, although in the cubic phase not all are occupied. The Li(1) atoms (8$a^{\rm t}$ and 24$d^{\rm c}$) are large gray (gold online), Li(2) or 16$f^{\rm t}$ are white, and Li(3) or 32$g^{\rm t}$ are dark gray. The cubic Li(1) positions that become vacant upon transition to the ordered tetragonal structure (16$e^{\rm t}$) are indicated by small (gold online) spheres.}\label{fig:cub_tetr}
\end{figure}

The Li sublattices in the cubic and tetragonal LLZO phases are shown in Fig.~\ref{fig:cub_tetr}. Li positions in each structure are generally referred to as Li(1) if they are tetrahedrally coordinated to oxygen, and Li(2) and Li(3) if they are octahedrally coordinated. To avoid confusion, we refer to each site using its crystallographic notation, with superscript $c$ or $t$ to designate cubic and tetragonal lattices, respectively. Each tetrahedral cubic 24$d^{\rm c}$ site is surrounded by four pairs of octahedrally coordinated 96$h^{\rm c}$ sites, and all sites are partially occupied (see Fig.~\ref{fig:cub_tetr}). Because of Coulomb repulsion, it is energetically prohibited for both members of each pair of 96$h^{\rm c}$ sites to be occupied, and if a particular 24$d^{\rm c}$ site is occupied, the adjacent 96$h^{\rm c}$ sites are consistently unoccupied \cite{Xie}. The tetragonal distortion transforms the cubic 24$d^{\rm c}$ sites into fully occupied 8$a^{\rm t}$ sites, often denoted as Li(1), and unoccupied 16$e^{\rm t}$ sites. The 96$h^{\rm c}$ sites are transformed into fully occupied 16$f^{\rm t}$ and 32$g^{\rm t}$ sites, often denoted as Li(2) and Li(3), respectively, with the rest fully unoccupied.  As we will show, a shift from 8$a^{\rm t}$ sites to 16$e^{\rm t}$ sites, both subsets of 24$d^{\rm c}$, always accompanies the tetragonal to cubic transition.

All of our simulations use density-functional theory, with the Perdew-Burke-Ernzerhoff exchange-correlation functional~\cite{Perdew:1996ug,Perdew:1997wu} as implemented in the Vienna Ab-initio Simulation Package (VASP) projector-augmented-wave software version 5.2.12~\cite{Kresse:1993ty,Kresse:1996vf}. The calculations used an energy cutoff of 400~eV, except for the relaxations used to explain the microscopic basis of the structural transition, which used a cutoff of 600~eV. Molecular dynamics (MD) time evolution is carried out using the velocity Verlet algorithm, as implemented in the libAtoms~\cite{libatoms_web} package, with a time step of 1~fs and using forces and stresses calculated at each time step with VASP. All MD simulations use constant temperature and stress. The time propagation algorithm, specified in Supplementary Information, is a rediscretization, using the ideas from Ref.~\onlinecite{Jones:2011fh} and Ref.~\onlinecite{Leimkuhler:2011ht}, of the Langevin constant pressure algorithm in Ref.~\onlinecite{Quigley:2004de}, for a modified version of the equations of motion of Ref.~\onlinecite{Tadmor:2011ue}. Each simulation starts from a unit cell of the experimental tetragonal structure of Ref.~\onlinecite{Awaka:2009jv}, for a total of 8 formula units. The structure is relaxed with respect to ionic positions and unit cell size and shape using VASP, and then simulated at finite temperature and zero stress.  While the MD simulation results implicitly include the effects of entropy, all energies we calculate explicitly and quote here include only DFT total internal energy.  The system is allowed to evolve for at least 24~ps with ionic motion as well as  cell shape and volume changes, so it can spontaneously transform into other structures. To determine the effect of composition and temperature we simulate the stoichiometric system, with 56 Li atoms, as well as systems with 1, 2, and 4 Li$^+$ ion vacancies per simulated cell, at temperatures ranging from 300~K to 1300~K. The net negative charge of the vacancies is compensated by a uniform positive background charge. The vacancies are formed by removing Li$^+$ ions randomly selected from the tetragonal 16$f^{\rm t}$ and 32$g^{\rm t}$ sites, where the vacancy formation energy is about 100 meV lower than at the 8$a^{\rm t}$ sites. The crystallographic site identity of each atom is determined during each MD trajectory (see Supplementary Information).

\begin{figure}
\centerline{\includegraphics[width=\columnwidth]{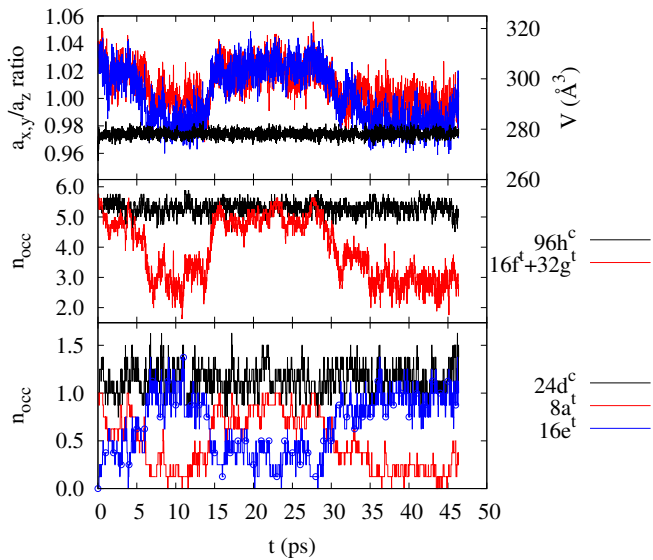}}
\caption{(Color online) Evolution over time of structural and site occupation quantities for a sample system with $n_{\rm vac}=2$ at $T=600$~K. Top: unit cell shape ($a_x/a_z$ blue, $a_y/a_z$ red) and volume (black). Middle: 96$h^{\rm c}$ (black) and 16$f^{\rm t}$+32$g^{\rm t}$ (red) lattice site occupation. Bottom: 24$d^{\rm c}$ (black), 8$a^{\rm t}$ (red) and 16$e^{\rm t}$ (blue with symbols) lattice site occupations. }\label{fig:time_evol}
\end{figure}

An example of the time evolution of one system, with $n_{\rm vac}=0.25$ and $T=600$~K, is plotted in Fig.~\ref{fig:time_evol}. The ratios of the lattice constants along $x$ and $y$ to that along $z$ initially show the tetragonal structure with one axis ($a_z$) about 3\% shorter than the others. The system transforms into a cubic phase where both axis ratios fluctuate around 1 at $t \sim 5$~ps. It fluctuates back to a tetragonal phase at $t \sim 15$~ps, and then again to cubic at $t \sim 30$~ps where it remains for the rest of the simulation. The volume is not affected by the unit cell shape change. We also plot the occupancies of various crystallographic sites of the cubic and tetragonal structures, computed as described above. We see a perfect correlation between the symmetry of the unit cell parameters and the occupancy of the tetragonal sites. The 8$a^{\rm t}$ occupancy is initially near 1, as it is in the experimental structure, and most of the remaining atoms are identified as 16$f^{\rm t}$ and 32$g^{\rm t}$. When projected into the cubic structure the 8$a^{\rm t}$ atoms are identified as 24$d^{\rm c}$ and the 16$f^{\rm t}$ and 32$g^{\rm t}$ atoms are identified as 96$h^{\rm c}$, as expected by symmetry. Whenever the system transforms into the tetragonal structure the occupancies of 8$a^{\rm t}$ and 16$f^{\rm t}$+32$g^{\rm t}$ sites drop. Since we do not see a corresponding change in 24$d^{\rm c}$ and 96$h^{\rm c}$, we conclude that the atoms are moving from the subset of the cubic sites that are occupied in the tetragonal ordering to the other cubic structure sites. This is indeed seen in the increased occupancy of 16$e^{\rm t}$ sites, which correspond to 24$d^{\rm c}$ sites that are unoccupied in the experimental tetragonal structure.

The unit cell shape, 8$a^{\rm t}$ occupancy, and unit cell volume, averaged over the last 5~ps of each run, are listed in Table~\ref{table:struct_avgs}. At each vacancy concentration the system transforms from an ordered tetragonal structure at low $T$ to a cubic structure at higher $T$, and the transition temperature goes down with increasing vacancy concentration. 
Note that at $n_{\rm vac}=0.5$ and $T=300$~K our simulations predict an ordered tetragonal structure that is different from the experimentally observed low $T$ tetragonal structure for the stoichiometric composition.  The structure we find has one long and two short axes, an occupation of the original tetragonal 8$a^{\rm t}$ sites of 0.5, and an occupation of 1 of the original tetragonal 16e$^{\rm t}$ sites (which are unoccupied in the stoichiometric tetragonal structure).
The volume is much more sensitive to temperature than to vacancy number, with a linear thermal expansion coefficient (from a linear fit of $V^{1/3}$ vs.\ $T$ to the $n_{\rm vac}=0.5$ results, where we have the widest range of temperatures) of $2.2 \times 10^{-4}$~K$^{-1}$. The vacancy number dependence (from a linear fit of $V$ vs.\ $n_{\rm vac}$ to the $T=600$~K and $T=800$~K results), which is the vacancy formation volume, is 10$\pm$1.5~{\AA}$^3$/vacancy. Note that this value is dependent on the charge compensation mechanism, which is a uniform background charge in this calculation, compared to any of a variety of supervalent dopants in experiment. In all cases, a transition to cubic symmetry is preceded by a sudden decrease in the 8$a^{\rm t}$ occupation and any return to tetragonal symmetry is characterized by a reoccupation of these sites.

\begin{table*}
\caption{Mean unit cell distortion ($a_\alpha/a_z-1$ for $\alpha$=$x$ and $y$, in \%) (top) and occupancy of 8$a^{\rm t}$ sites per formula unit (bottom), averaged over the last 5~ps of each run, for different vacancy numbers $n_{\rm vac}$ and temperatures. Transition temperature $T_c$ is also indicated for each quantity with a clear signal of a structural transition (see text). 
}\label{table:struct_avgs}
\begin{ruledtabular}
\begin{tabular}{lccccccc}
$n_{\rm vac}$ & 300~K     & 450~K     & 600~K     & 800~K     & 1000~K    & 1300~K &   $T_c$ \\ 
\colrule
\multicolumn{7}{c}{$a_x/a_z$, $a_y/a_z$} \\
0.00 &      &      & 3.2, 3.3 & 2.3, 2.8 & 0.3, $-$0.6 & $-$0.1, $-$0.2 & 800~K$\le T_c \le$1000~K \\
0.12 &      &      & 3.0, 2.7 & 1.7, 2.1 & $-$0.2, $-$1.8 &       & 800~K$\le T_c \le$1000~K \\
0.25 & 3.3, 3.2 & 0.5, $-$2.2 & $-$0.5, $-$1.6 & $-$0.6, 0.7 &     &       & 300~K$\le T_c \le$450~K \\
0.50 & 1.3, $-$0.3 & 0.8, 0.3 & $-$0.0, $-$0.4 & $-$0.2, $-$0.5 & $-$0.2, $-$0.4 & 0.3, 0.0 & $T_c \le $300~K \\
\\
\multicolumn{7}{c}{8$a^{\rm t}$ occupancy} \\
0.00 &   &   & 1.0 & 0.9 & 0.3 & 0.3 & 800~K$\le T_c \le$1000~K \\
0.12 &   &   & 0.9 & 0.8 & 0.2 &   & 800~K$\le T_c \le$1000~K \\
0.25 & 1.0 & 0.2 & 0.2 & 0.3 &   &   & 300~K$\le T_c \le$450~K \\
0.50 & 0.5 & 0.4 & 0.4 & 0.3 & 0.3 & 0.4 & ? \\
\end{tabular}
\end{ruledtabular}
\end{table*}

\begin{figure}
\centerline{\includegraphics[width=\columnwidth]{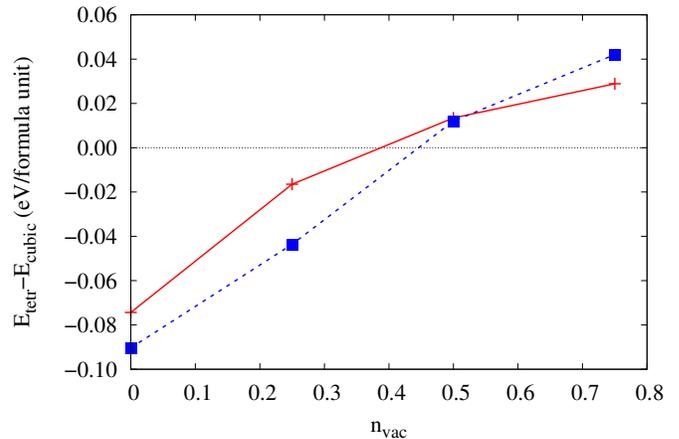}}
\caption{(Color online) Energy difference between tetragonal and cubic structures as a function of vacancy number, for minimum energy configuration (solid red line) and mean configuration energy (dashed blue line).}\label{fig:tetr_cubic_diff}
\end{figure}

We have also computed the energy difference between the tetragonal and cubic structures as a function of vacancy concentration. For the tetragonal structure we relaxed 10 structures with random vacancies at 16$f^{\rm t}$ and 32$g^{\rm t}$ sites. For the cubic structure, which is disordered even at the stoichiometric composition, we use 10 configurations from a uniform sampling of all configurations that obey the restriction that no pairs of nearest neighbors (adjacent 96$h^{\rm c}$ pairs, or a 24$d^{\rm c}$ and its nearest neighbor 96$h^{\rm c}$) are simultaneously occupied. We plot the energy difference between the minima and means of each of the ten structures in Fig.~\ref{fig:tetr_cubic_diff}. The tetragonal structure is energetically favored for $n_{\rm vac} \le 0.25$, and the cubic is favored for $n_{\rm vac} \ge 0.5$. 
Entropy effects will shift the equivalent curves for the free energies.  We expect that this shift will reduce the free energy of the tetragonal phase more than the cubic phase, because the tetragonal phase starts out ordered and therefore gains more entropy with the introduction of vacancies than the already disordered cubic phase.  The transition vacancy concentration will therefore shift to higher values as the temperature increases, although we expect this shift to be small at $T=300$~K.

\begin{figure}
\centerline{\includegraphics[width=0.9\columnwidth]{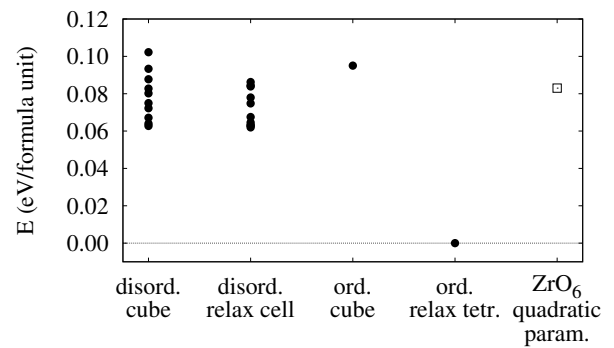}}
\caption{Relaxed energies (closed symbols) of 10 disordered configurations and one at the experimentally observed tetragonal phase order, for a cubic unit cell and for a fully relaxed one, relative to the lowest energy fully relaxed ordered tetragonal cell. The energy of a model including only ZrO$_6$ distortions is also plotted (open symbol).}\label{fig:energy_spectrum}
\end{figure}

To understand the relationship between Li order/disorder and the tetragonal/cubic lattice parameters, we perform total energy calculations for the stoichiometric system in 10 disordered configurations (again from a uniform sampling of configurations obeying first-neighbor exclusion) and in the experimentally observed Li order in the tetragonal phase. We first relax each one with respect to atomic positions and unit cell volume but constrain the unit cell shape to remain cubic, and then relax with full freedom for the unit cell size and shape as well as atomic positions. The total energies are plotted in Fig.~\ref{fig:energy_spectrum}. The energies of the disordered system are generally highest, with only a small gain (and correspondingly small change in unit cell shape) from relaxing the cubic unit cell constraint. Imposing the Li order while maintaining a cubic unit cell actually increases the energy slightly as compared with the mean disordered energy. Only once the ordered cell is allowed to relax to the tetragonal shape does its energy become lower than the disordered system, and the ordered system emerges as the $T=0$~K ground state.

To understand the source of the coupling between energy and structure, we calculated the pair distribution functions (PDF) of various ion pairs in the system for the cubic and tetragonal ordered systems, along with a simple point charge model, based on the nominal ionic charges. The point charge model shows that the overall Coulombic energy is actually {\em increased} upon relaxation to a tetragonal unit cell, indicating that the energy gain is elsewhere. The PDFs show that the La$^{3+}$$-$O$^{2-}$ distances do not vary, presumably since these are the two largest ionic charges in the system. Li$-$O and Li$-$Li distances do change, but the highly ionic nature of these interactions means that such changes are nearly entirely accounted for in the point charge model. The shifting of Li$-$Li and Li$-$O distances, combined with the rigidity of the La$-$O spacing, results in some distortion of the ZrO$_6$ octahedra. Unlike the other pairs, the Zr$-$O interaction is at least partially covalent. In the cubic cell, the Zr$-$O bond lengths are 2.130$\pm$0.020 {\AA} and O$-$Zr$-$O bond angles are 180$\pm$4.0$^\circ$. The relaxation to a tetragonal cell relieves this distortion, restoring the octahedra to a uniform Zr$-$O bond length of 2.125$\pm$0.005~{\AA} and a O$-$Zr$-$O bond angle of 180$\pm$0.01$^\circ$. To estimate the energetic contribution of ZrO$_6$ octahedra distortions we parametrize the calculated total energies of the ordered tetragonal cell as a quadratic function of Zr$-$O bond length and O$-$Zr$-$O bond angle by computing the energies for small displacements of an O atom. This parametrization predicts an energy for the ordered cubic structure of 0.083~eV/formula unit relative to the tetragonal structure (Fig.~\ref{fig:energy_spectrum}), very nearly equal to the DFT energy difference. We therefore conclude that lithium ordering to relieve Li-Li Coulomb repulsion leads to internal rearrangements of ions to maintain La$^{3+}$$-$O$^{2+}$ distances, which leads to a lattice distortion that allows the ZrO$_6$ octahedra to preserve their preferred shape.

In summary, our variable cell shape DFT MD simulations of LLZO show that the DFT ground state at low temperatures is the experimental ordered tetragonal structure, and that at higher temperatures the system transforms into the experimental disordered cubic structure. The transition temperature decreases with increasing Li$^+$ vacancy concentration, and the disordered cubic structure has lower energy when the number of vacancies per formula unit is larger than about 0.4. These results are in agreement with the experimental observations of a transition as a function of Li$^+$ vacancy concentration~\cite{Allen:2012gk}. The microscopic cause of the structural transition is the coupling between the unit cell shape and the hopping of atoms from the subset of the disordered cubic sites occupied in the ordered tetragonal structure to the remaining cubic sites, which are unoccupied in the tetragonal structure. The relative stability of the ordered tetragonal low temperature structure is driven by the ordering, which reduces Li$-$Li Coulomb repulsion but distorts the ZrO$_6$ octahedra, and the tetragonal distortion which allows these octahedra to return to their preferred high-symmetry shape. Our simulations enable us to explain the atomistic mechanism behind this finite temperature structural transformation in a complex material, and predict the number of vacancies necessary to achieve the high conductivity material. This should enable better doping schemes that optimize ionic conductivity by providing the requisite number of vacancies with as few artificial dopants as possible, thereby realizing the potential of LLZO as a truly stable solid electrolyte material.

This work was supported by the Naval Research Laboratory core 6.1 research program
and the Nanoscience Institute.  Computer time was provided through the DOD HPCMPO
at the ERDC and AFRL MSRCs.

\bibliography{LLZO_var_cell_shape_Papers,LLZO_var_cell_shape_MDJ,LLZO_var_cell_shape_extra}

\end{document}